\documentclass[aps,prl,twocolumn,showpacs,superscriptaddress,amssymb]{revtex4}
\usepackage{graphicx}
\usepackage{hyperref}
\usepackage{natbib}
\usepackage{bm}
\begin{document}

\title{Decay of Rabi oscillations by dipolar-coupled dynamical spin
environments}

\author{V. V. Dobrovitski}
\affiliation{Ames Laboratory, Iowa State University, Ames IA 50011, USA}
\author{A. E. Feiguin}
\affiliation{Condensed Matter Theory Center, 
Department of Physics, University of Maryland, College Park, MD 20742}
\affiliation{Microsoft Station Q, University of California, Santa Barbara, 
CA 93106.}
\author{R. Hanson}
\affiliation{Kavli Institute of Nanoscience Delft, Delft University of Technology, 
P.O. Box 5046, 2600 GA Delft, The Netherlands}
\author{D. D. Awschalom}
\affiliation{Center for Spintronics and Quantum Computation, University 
of California, Santa Barbara, California 93106, USA}

\date{\today}

\begin{abstract}
We investigate the Rabi oscillations decay
of a spin decohered by a spin bath
whose internal dynamics is caused by dipolar coupling between the bath
spins. The decay form and rate as a function of the
intra-bath coupling is studied analytically, and confirmed 
numerically. The decay in general has neither 
exponential/Gaussian or power-law form, and
changes non-monotonically with the intra-bath coupling, decelerating
for both slow and fast baths.
The form and rate of Rabi oscillations decay can 
be used to experimentally
determine the intra-bath coupling strength for a broad class of 
solid-state systems.
\end{abstract}


\maketitle

Measurement of the Rabi oscillations decay is an important step
in studying decoherence of quantum systems. For instance, extensive
studies of Rabi oscillations in superconducting qubits 
\cite{RabiJJ2,RabiJJ3,RabiJJ4}
greatly enhanced understanding of the decoherence caused by
the bosonic baths and by the $1/f$ noise. The
most promising directions for improvement were identified, and 
the decoherence time for superconducting qubits
have been extended to microseconds range \cite{RabiJJ5}, placing
them among most promising solid-state qubit candidates.
Recently, much progress has been achieved in experimental implementation
of long-living coherent Rabi oscillations in 
various 
spin systems: magnetic molecules \cite{RabiV15}, NV 
impurity centers
in diamond \cite{RabiNV1,RabiNV2,RabiNV3}, rare-earth ions in non-magnetic host 
\cite{RabiRE}, quantum dots \cite{RabiQD1,RabiQD2,RabiQD3}, to mention a few.
Many of these systems are very attractive for basic studies of
quantum spin coherence effects, and also constitute promising candidates for
quantum information processing, coherent spintronics, or high-precision
magnetometry devices, provided that detailed understanding of the decoherence 
processes will be achieved. 

A major decoherence
source in these spin systems is the coupling of the central spin 
(e.g.\ the electron
spin of the NV center in diamond) to other spins 
present in the sample (environmental bath spins, e.g.\ the spins of nitrogen 
atoms in diamond). Moreover, for many relevant spin 
systems, the direct coupling between the environmental spins is essential,
producing internal dynamics of the bath.
Here we explore theoretically
the influence of such a dynamical spin bath on the Rabi oscillations of the
central spin. Avoiding the commonly used framework of generalized Bloch 
equations \cite{Slichter,Abragam,CohTan,Kosloff}, we are able to investigate 
the form and rate of decay 
as a function of the intra-bath coupling strength.
We find interesting behavior of the Rabi oscillations,
which contradicts the expectations based
on standard Redfield-type analysis: e.g.,
the slow bath leads to pronounced decay, while for the fast bath
the decay rate vanishes but the Rabi frequency becomes renormalized.
We demonstrate how 
the form and rate of the Rabi oscillations
decay can be used in experiment to characterize
the intra-bath dynamics, and provide a rather simple 
recipe for analysis of experimental data.

Specifically, we consider the situation of dilute dipolar-coupled bath: 
in a non-magnetic host crystal, a single central
spin of species $S$ (e.g.\ belonging to a paramagnetic impurity) 
is coupled to a bath of dilute spins of species $I$ (e.g.\ 
belonging to other kind of
paramagnetic impurities or to nuclear spins). The coupling
between the bath spins is non-negligible, and is caused by
dipolar interactions. This situation encompasses
a wide range of interesting solid-state spin systems,
from Er ions in CaWO$_4$ studied in 1960s \cite{Mims1968} to
the NV centers in diamond \cite{RabiNV1,RabiNV2,RabiNV3} which
gained much attention very recently. 
Moreover, experimental progress now makes possible detailed studies
of Rabi oscillations of an individual $S$ spin \cite{RabiNV1,RabiNV3,MRFM}.
We assume that a large 
magnetic field $B_z$ is applied along the $z$-axis (as in standard
NMR/ESR settings), leading to Zeeman splittings $\omega_S=\gamma_S B_z$
for the central spin $S$ and $\omega_I=\gamma_I B_z$ for the bath spins $I_k$
($\gamma_S$ and $\gamma_I$ are the gyromagnetic ratios of the $S$ and $I$ spins,
respectively). Also, strong Rabi driving field $H_R$ is applied at the
frequency $\omega_S$. The difference $|\omega_S-\omega_I|$ is much larger than
any other energy scale, and after transformation into rotating frame
we obtain a standard secular Hamiltonian for two dipolar-coupled spin species
\cite{Slichter,Abragam}:
\begin{equation}
\label{HamInit}
H= h_x S_x + \sum_k A_k S_z I_k^z + \sum_{k,l} C_{kl}(3 I_k^z I_l^z-{\mathbf I}_k {\mathbf I}_l)
\end{equation}
where $S^{x,y,z}$ and $I^{x,y,z}_k$ are the spin operators in the rotating frame,
and $h_x=H_R/2$ is the rotating-frame Rabi driving field.
The 
coupling constants $A_k=\gamma_S\gamma_I [1-3(n^z_k)^2]/r_k^3$ 
are determined
by the positions ${\mathbf r}_k$ of the bath spins $I_k$ ($k=1,\dots N$), where 
$r_k=|{\mathbf r}_k|$ and ${\mathbf n}_k={\mathbf r}_k/r_k$ 
(the origin of the coordinate frame coincides with the central spin).
Similarly, the intra-bath couplings $C_{kl}=\gamma^2_I [1-3(n^z_{kl})^2]/r_{kl}^3$
are determined by the vectors
${\mathbf r}_{kl}={\mathbf r}_k - {\mathbf r}_l$. Note
that the same Hamiltonian (\ref{HamInit}) can be obtained without external field
$B_z$ if the transition frequencies $\omega_S$ and $\omega_I$ are determined by
the zero-field splitting (e.g.\ due to anisotropic interactions). Similarly,
the assumption (used below) that $S$ and $I_k$ are spins 1/2 is not essential:
for larger spins, we can consider each pair of levels 
as a pseudo-spin 1/2 \cite{Slichter,Abragam}.

Initially the central spin is in the state "up", and
the bath is in a maximally mixed state (unpolarized
bath at high temperatures), i.e. the initial density matrix of the
whole system is $\rho(0)=2^{-N} |\uparrow\rangle\langle\uparrow|\otimes{\mathbf 1}_B$,
where ${\mathbf 1}_B$ is the $2^N\times 2^N$ identity matrix. 
This is appropriate for most experiments 
(for nuclear spins at temperatures above few nK, for electron spins
--- above tens of K). We calculate the time-dependent 
$z$-component of the central spin $\langle S_z(t)\rangle={\rm Tr} \rho(t)S_z$,
where $\rho(t)=\exp{(-iHt)}\rho(0)\exp{(iHt)}$.
In order to see well-pronounced long-living Rabi oscillations, the
driving field should be large, so we assume that $h_x$ is much larger
than all other energy scales. We calculate the oscillations damping in
the lowest order in $1/h_x$, treating the second (spin-bath coupling)
and the third (bath internal Hamiltonian) terms in Eq.~\ref{HamInit}
perturbatively,
and excluding the bath internal Hamiltonian 
$H_B=\sum_{k,l} B_{kl}(3 I_k^z I_l^z-{\mathbf I}_k {\mathbf I}_l)$
by the interaction representation transformation. The resulting
second-order
Hamiltonian 
\begin{equation}
\label{Ham0}
H_0= h_x S_x + S_x {\hat B}^2(t)/(2 h_x)
\end{equation}
where ${\hat B}(t)=\exp{(i H_B t)} {\hat B} \exp{(-i H_B t)}$, and 
the operator ${\hat B}=\sum_k A_k I_k^z$.

The evolution of ${\hat B}(t)$ is complex, involving intricate correlations
between bath spins. However, exact dynamics of every single bath spin is not important,
since $\langle S_z(t)\rangle$ involves tracing 
over all bath spins.
This is typical for many spin-bath decoherence problems,
from magnetic resonance to quantum information processing, 
and many approaches 
have been developed from the early
days of NMR/ESR theory \cite{PWA,Kubo} till very recently
\cite{Garg,VVD,coish,dassarma,sqrtTh,cucch,taylor,lusham}.
All approaches involve
a trade-off between quantitative rigour and qualitative understanding.
Below, following the works \cite{PWA,Kubo}, we approximate the
effect of the bath by a random field $B(t)$, which is modeled
as an Ornstein-Uhlenbek process with the correlation function
$\langle B(t) B(0)\rangle = b^2 \exp{(-Rt)}$, where the
dispersion $b=\sqrt{\sum_k A_k^2}$, while the correlation decay rate $R$
is determined by $H_B$. 
This model
may be oversimplified for complicated situations, e.g.\ the description
of advanced control protocols requires more sophisticated treatment 
\cite{dassarma,lusham}. 
However, our direct numerical simulations
evidence that this model is {\it quantitatively\/} adequate for
description of the Rabi oscillations decay, while
providing a clear description of the physics underlying the
Rabi oscillations decay, and allowing access to the regimes outside of
the Bloch equations framework.

The Hamiltonian (\ref{Ham0}) reflects a simple physical picture.
The zero-order eigenstates 
of the Hamiltonian (\ref{HamInit}) correspond to the central spin states 
$|+\rangle=1/\sqrt{2}[|\uparrow\rangle + |\downarrow\rangle]$ and
$|-\rangle=1/\sqrt{2}[|\uparrow\rangle - |\downarrow\rangle]$, separated
by a large Rabi frequency $h_x$. Since $h_x\gg b$ and $h_x\gg R$, the field
$B(t)$ has no spectral components of noticeable magnitude at the 
Rabi frequency, and does not lead to transition between the states
$|+\rangle$ and $|-\rangle$. The only relevant process 
is the pure dephasing, when the field $B$
destroys the initial phase relation between the states $|+\rangle$ and $|-\rangle$,
leading to decay of $\langle S_z(t)\rangle$, so that
\begin{equation}
\langle S_z(t)\rangle = 1/2\ \langle\cos{\Phi}\rangle = 
  1/2\ {\rm Re}\ \langle\exp{(i\Phi)}\rangle
\end{equation}
where $\langle\cdot\rangle$ denotes average over all possible realizations
of $B(t)$, and the phase
\begin{equation}
\Phi = h_x t + \frac{1}{2h_x}\int_0^t B^2(s) ds = h_x t + \Theta
\end{equation}
is the total phase difference
between the states $|+\rangle$ and $|-\rangle$ accumulated during time $t$,
cf.\ the Hamiltonian (\ref{Ham0}). The key quantity 
$M(t)=\langle\exp{(i\Theta)}\rangle$
is an analytically computable
Gaussian path integral over the Ornstein-Uhlenbeck process, which gives the 
answer
\begin{eqnarray}
\label{sol}
\nonumber
\langle S_z(t)\rangle &=& 1/2\ {\rm Re} \left[M(t)\exp{(ih_xt)}\right]\\
\nonumber
[M(t)]^{-2} &=& \exp{(-Rt)}\left[\cosh{Pt} + (R/P)\sinh{Pt}\right] \\ 
    &-& i\frac{b^2}{h_x P} \exp{(-Rt)}\sinh{Pt},
\end{eqnarray}
where $P=\sqrt{R^2-2i b^2 R/h_x}$.

Analysis of Rabi oscillations is often based on Bloch-type
equations \cite{Slichter} or its generalizations \cite{Kosloff},
derived for various systems from quantum optics \cite{CohTan}
to solid state \cite{Kosloff}. It is based on the Redfield-type approach
\cite{Slichter,Abragam},
taking into account only the terms which are secular with respect 
to the Rabi driving $h_x S_x$.
In addition to dephasing, these terms describe 
the actual transitions between the states $|+\rangle$ and $|-\rangle$, which
lead to a longitudinal relaxation (along the $x$-axis) of the central spin
with the rate $\Gamma_l\sim b^2 R/h_x^2$. This rate
is of second order in $1/h_x$, and is determined by the spectral density of 
$B(t)$ at the Rabi frequency $h_x$. Also, the generalized Bloch
equations, having constant coefficients, always predict
the 
decay to have a (multi)exponential form. 
In contrast, our results are not 
limited to the terms secular
with respect to Rabi driving, and give the decay rate of the first
order in $1/h_x$. 
The solution (\ref{sol}) predicts the decay 
which has, in general, no simple form (multiexponential, Gaussian, power-law, etc.)
The familiar exponential 
decay occurs
only at special values of $b$, $h_x$, and $R$, see below. 

The effect of the bath internal dynamics may be important even for
slow baths, with $R\ll b^2/h_x$. In this limit,
$P=b\sqrt{-2iR/h_x}$, see Eq.~\ref{sol}. At short times $t\ll 1/|P|$, 
the bath behaves as static, and the Rabi oscillations
envelope exhibits non-exponential slow decay of the form 
$\left[1+(b^2 t/h_x)^2\right]^{-1/4}$,
in accordance with the exact results obtained earlier
\cite{sqrtTh,cucch,taylor,RabiQD1}. At long times $t\gg 1/|P|$, 
the decay has exponential form $\exp{(-bt\sqrt{R/4h_x})}$, 
with the rate which decreases very slowly with decreasing $R$. 
As a result, even for very slow bath the effect of the
bath dynamics remains noticeable. In experiments,
the above-described behavior of the Rabi oscillations decay
can be detected by varying $h_x$ (since the ratio of $R$ to $b^2/h_x$
determines how fast/slow the bath is), and makes it possible to estimate 
$b$ and $R$.
%

\begin{figure}
\includegraphics[width=8cm]{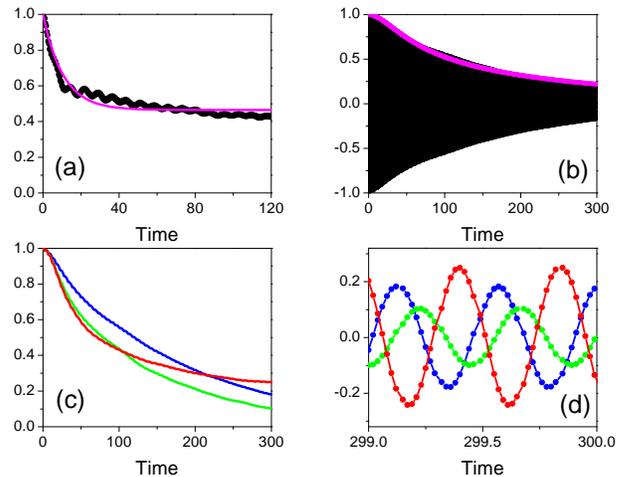}
\caption{(Color online).
Exact numerical simulations for $N=15$ bath spins.
(a) --- Correlation function $\langle B(0)B(t)\rangle$ (normalized
to 1 at $t=0$) obtained
from direct simulations (black), see also \cite{noteSmallN}. Its fitting (green) 
determines the bath parameters $b_1$, $b_2$, and $R$ for each value of the intra-bath
coupling scale $E_B$ (here $E_B=1$,
$b_1=0.62$, $b_2=0.58$, $R=0.095$). (b) --- Rabi oscillations decay
for $h_x=14.14$, $b=0.85$, and $E_B=1$; individual oscillations are
not resolved. Numerical results (black) agree well with analytics 
(magenta line, only the envelope shown). (c) --- Envelopes of simulated
Rabi decay for $E_B=1$ (blue), $E_B=0.1$ (green), and $E_B=0$ (static bath,
red); corresponding individual oscillations near $t=300$ are shown on panel (d)
by same colors. Analytical
results practically coincide with simulations, and are not shown.
The decay rate changes nonmonotonically with $E_B$: the decay is slower 
for $E_B=1$ (blue), and $E_B=0$ (red) in comparison with $E_B=0.1$ (green).
At $t=300$, on panel (d), the oscillation amplitude for $E_B=0.1$ (green) is
twice smaller than that for $E_B=0$.
\label{figN15}}
\end{figure}

Another interesting feature of our results is that the decay
of Rabi oscillations changes non-monotonously with $R$: it is
fastest for $R\sim b^2/h_x$, slowing down for both slow and fast baths.
This is confirmed by direct numerical simulations, see Fig.~\ref{figN15}
and discussion below. 

\begin{figure}
\includegraphics[width=8cm]{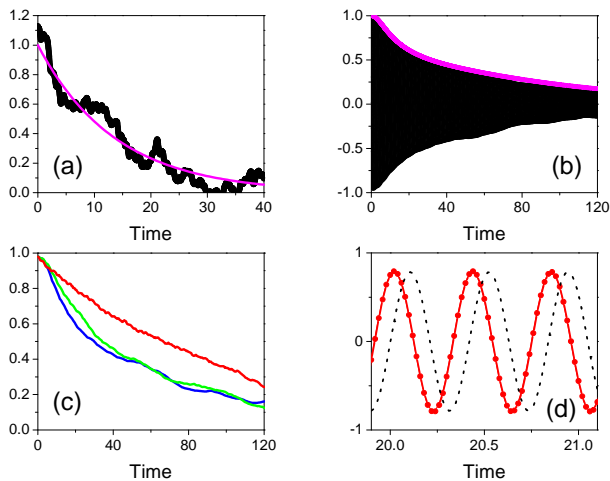}
\caption{(Color online).
Numerical simulations for $N=159$ bath spins.
(a) --- Correlation function $\langle B(0)B(t)\rangle$ (normalized
to 1 at $t=0$) obtained
from P-representation simulations (black); its fitting (green) 
determines the parameter $R$ for each value of the intra-bath
coupling scale $E_B$ (here $E_B=1$ and $R=0.097$). In spite of rather large
statistical fluctuations, the
parameter $R$ is determined precisely enough to be used in Eq.~\ref{sol}. 
(b) --- Rabi oscillations decay for $h_x=15.0$, $b=1.39$, and $E_B=0.1$;
individual oscillations are not resolved. Numerical results (black) 
agree well with analytics 
(magenta line, only the envelope shown). (c) --- Envelopes of simulated
Rabi decay for $E_B=0.1$ (blue), $E_B=1$ (green), and $E_B=10$ (red);
the longitudinal decay for $E_B=10$ is taken into account. Analytical
results practically coincide with simulations, and are not shown.
The decay rate decreases for faster baths: the decay
is slowest for $E_B=10$ (red). Individual oscillations for $E_B=10$
near $t=20$ are shown on panel (d) in red: their frequency 
is shifted by $b^2/2h_x=0.064$ from
the value $h_x=15.0$ (oscillations with frequency $h_x=15.0$
are shown by dotted black line to demonstrate the phase difference accumulated
since $t=0$).
\label{figN159}}
\end{figure}

The regime of fast bath, with $R\gg b^2/h_x$, is even more interesting 
and unexpected. In this case
\begin{equation}
M(t)=\exp{\left[ i\frac {b^2 t}{2 h_x} - \frac{b^4 t}{4 h_x^2 R}\right]},
\end{equation}
and the Rabi oscillations exhibit exponential decay with the 
rate $\Gamma_t=b^4/(4 h_x^2 R)$, which
vanishes quickly for large $R$. However, it does not mean that the effect
of the fast bath disappears. The bath still noticeably affects the central
spin, shifting its Rabi frequency by $b^2/2 h_x$, and only
the decaying part of the bath effect vanishes \cite{MotNarNote,GenNote}. 
This slowing of the Rabi oscillations decay is confirmed by direct numerical 
simulations (Fig.~\ref{figN159}). 

On the other hand, for fast baths, the transitions between the
states $|+\rangle$ and $|-\rangle$ become important.
While the dephasing between these states just shifts the Rabi frequency,
the transitions lead to longitudinal relaxation along the $x$-axis
with the rate $\Gamma_l\sim b^2 R/h_x^2$. This
implies \cite{Slichter}
exponential decay of Rabi oscillations with the rate $\Gamma_l/2$,
which is comparable with the decay rate $\Gamma_t$ caused by the pure dephasing.
Above, for simplicity, we omitted the longitudinal relaxation
(since it is of order $1/h_x^2$), but we can include it using 
the Redfield-type approach:
the answer (\ref{sol}) for $S_z(t)$ just has to be multiplied by
$\exp{(-\Gamma_l t/2)}$. This explains how the system enters the generalized
Bloch equation regime at $R\gg b$: the dephasing effect becomes
negligible and the longitudinal relaxation becomes dominant.
In experiment, this regime can be identified by comparing $\Gamma_l$
and $\Gamma_t$: they would be of the same order, changing as $1/h_x^2$ for 
all sufficiently large $h_x$.

To ensure that the physical picture above is adequate
for real dipolar-coupled bath, 
we performed direct numerical simulations of the Rabi oscillations decay.
We place the central spin and $N$ bath spins randomly, 
with uniform density $n=1$, inside a cube with the side $(N+1)^{1/3}$,
central spin being in the center of the cube. All interaction coefficients
are calculated according to Eq.~\ref{HamInit}, using the actual
coordinates of the spins. The value 
$E_{SB}=\gamma_S\gamma_I$, which determines the strength of coupling
between the central spin and the bath, is set to 1, thus defining
the energy and time scales for all 
quantities below. 
The value $E_{B}=\gamma_I^2$, which governs
the energy scale of intra-bath couplings, is varied, making the bath slower or faster.
We simulate the dynamics of the system using two numerical approaches.
For small $N$ (e.g., $N=15$ in Fig.~\ref{figN15}) we exactly solve
of the Schr\"odinger equation with the Hamiltonian (\ref{HamInit}) via
Chebyshev polynomial expansion \cite{revVVD}. The 
dipolar-coupled systems with $N>15$ are difficult to model this way (due 
to exponentially increasing resources requirements), so we use
the P-representation sampling \cite{revVVD} for modeling larger baths 
($N=159$ in Fig.~\ref{figN159}).
%
%
To compare numerical solutions with the analytics,
we calculate the total coupling
to a bath 
$b=[{\sum_{k=1}^N A_k^2}]^{1/2}$ directly from
the positions of the spins (see Eq.~\ref{HamInit} and below).
The correlation decay rate $R$ requires
a separate simulation: we calculate
$\langle B(0)B(t)\rangle$ and find $R$ by fitting it to a decaying exponent,
see also \cite{noteSmallN}.
Varying the system parameters in a wide range, 
we simulated baths with $N=15$, 59, and 159 spins, 
and found good agreement between numerics and analytics;
typical results are given in Figs.~\ref{figN15},\ref{figN159}.

The resulting experimental recipe is rather simple. If the decay has a power-law
form at shorter times, changing to exponent later, with the decay constants
changing as $1/h_x$ and the duration of the two regimes varying with $h_x$,
then the bath is slow. If the Rabi oscillations decay is exponential,
with the rate changing as $1/h_x^2$ and the frequency shift varying as $1/h_x$,
then the bath is fast. The fast bath can be made slow by strong increase
in $h_x$.

Summarizing, we studied the decay of the Rabi oscillations of the central spin
interacting with a dipolar-coupled dynamical spin bath. 
Approximating the
effect of the bath as a random field (Ornstein-Uhlenbeck process),
we find analytically the form of the decay. Validity of the approximation
is confirmed by direct numerical simulations. The oscillations decay
has interesting features, such as 
non-monotonic variation of the decay rate with increasing the intra-bath
coupling, and slowing down of the decay for fast baths.
Studying the behavior of the Rabi oscillations may
help in experimental characterization of the dynamical spin bath,
and is well within experimental reach e.g.\ for NV centers in diamond.
Work at Ames Laboratory was supported by the Department of Energy 
--- Basic Sciences under Contract No.~DE-AC02-07CH11358.
We acknowledge support from AFOSR (D.D.A.), FOM and NWO (R.H.). 
A.E.F. 
acknowledges support from the Microsoft Corporation.


\begin{thebibliography}{99}
\bibitem{GenNote} This result can be obtained rather universally, for any non-pathological
bath, starting from Eq.~\ref{Ham0} and assuming
that the correlations $\langle B(0)B(t)\rangle$ decay to
zero with a rate much faster than $b^2/h_x$.
\bibitem{noteSmallN} For $N=15$, due to 
finite-$N$ effects, the correlations decay to a finite value, so
that the field $B(t)$, in addition to the Ornstein-Uhlenbeck component,
has a static component (of order of $1/\sqrt{N}$), and the correlation function 
has the form $\langle B(0)B(t)\rangle=b_1^2\exp{(-Rt)}+b_2^2$, with
restriction $b_1^2+b_2^2=b^2$. Simulations still agree with analytics very well.
\end{thebibliography}
\end{document}